\documentclass[11pt]{article}
\usepackage{graphicx}

\title{Basic considerations about experimental approaches to the $B$--mode of the CMB Polarization}

\author{S. Cortiglioni$^{1}$ and 
   E. Carretti$^{2}$\\
\\
\small $^{1}$INAF--IASF Bologna, Via Gobetti 101, I-40129 Bologna, Italy\\
\small $^{2}$INAF--IRA, Via Gobetti 101, I-40129 Bologna, Italy
}

\begin{document}

\maketitle

   \abstract{The $B$-mode detection of Cosmic Microwave Background polarization will require new technological developments, able to get sensitivities at least 2 orders of magnitude better than for the $E$-mode. This really ambitious goal cannot be reached simply by either improving the present technology or by adding more detectors to current design, at least in the frame of having a new space mission operating within a decade. Thus, the scientific community have to take important decisions about the most suitable technologies on which converge the needed effort. Basically, at present two receiver families do exist: bolometric and radiometric. Both of them are continuoulsly improving their basic performances, but the optimal approach to $B$-modes may require some decisions have to be taken in short time scale. In any case, radiometric and bolometric receivers have to deal with some common sources of systematics as well as they both require some cryogenics. Thus, we should expect that systematics and cryogenics may play as watershed line in designing future experiments aimed at measuring the $B$-mode of Cosmic Microwave Background.}

\section{Introduction}
The huge importance of investigating the Early Universe through measurements of Cosmic
Microwave Background Polarization (CMBP) is widely accepted by the astrophysical 
scientific community. By measuring the angular power spectra (APS) of the CMB it is possible
to study effectively the inflation era, scalar (density) and tensorial (gravitation waves)
primordial perturbations, formation processes of first stars and galaxies \cite{zss97}\cite{kinney99}\cite{cen03}. Their implications in High Energy Physics 
can be 
studied as well.

Despite of their challenging nature, at least with presently available technology, measurements of CMBP open up the strong possibility of providing truly remarkable new insights into Cosmology and Fundamental Physics. 
The importance of tools such the $E$-mode and $B$-mode APS have been discussed by other authors, even in this volume, and these aspects will not be within the arguments of this article. Even the current status of CMBP experiments will not be reviewed here, but the starting point shall be rather the overall picture provided by their results as they can be read on the literature. 

A plethora of experiments attempting to measure the $E$-mode APS are already operating or 
in progress and we should assume that they will get successfully to the aim. Updated 
references can be found in \cite{montroy05}, 
which reports the most recent result 
in the field. An almost complete list of suborbital experiments aiming at CMB investigations can be found in the web site http://lambda.gsfc.nasa.gov/product/suborbit/su\_experiments.cfm. 

However, the full characterization of the $E$-mode APS may be still beyond the capabilities of these experiments, which are constrained by their intrinsic limits to make only statistic detections at some angular scales. A first significant result for the $E$-mode APS full characterization is expected from the Planck ESA-satellite after 2007, unlikely before.

As in the case of CMB anisotropy (and black-body spectrum), which mesasurements began few decades ago, the early experimental approaches are done by using instruments derived from previous generation. The well known history, in fact, is that first attempts to detect anisotropy have been done with radioastronomical receivers, before understanding that new design and techniques were fundamental requirements to get to the end successfully. This was due mainly to the difficulties to remove both systematics and foregrounds at the required level, which is sligthly  different in radioastronomy.

For the CMBP the situation looks similar: present CMBP experiments, in fact, are based on receivers previously designed for measuring the anisotropy, even though using state-of-the-art components. Even the foreground removal techniques were based, initially, on data taken in total intensity and the knowledge of polarized foregrounds is limited to surveys done at frequencies really far from the 70-100 GHz window, that looks suitable for CMBP observations. More recently the study of polarized galactic emission has been boosted by observing in polarization some low emission area, from which it has been possible to set important implications on CMBP measurements \cite{CPR06} \cite{CBC06}. Thus, we should expect that a significant step in the full characterization of CMBP $E$-mode shall require: 
\begin{itemize}

\item{}	Dedicated instrument design
\item{}	Dedicated observing strategies
\item{}	Dedicated foreground removal techniques
\end{itemize}

\begin{figure}
 \centering
  \includegraphics[angle=0, width=10cm]{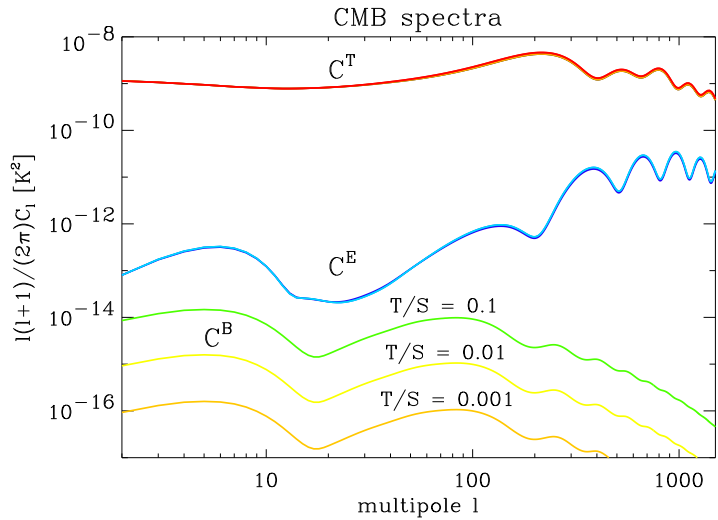}
  \caption{The CMB angular power spectra: anisotropy (red), $E$-mode (blue) and $B$-mode polarization (others). The latter is plotted for different T/S values.} 
  \label{cbFig}
\end{figure}

The way to measure the $B$-mode and its APS, which is a factor 10-40 tinier 
(Fig.~\ref{cbFig}) than for $E$-mode, looks definitely harder. At the moment 
no experiments aimed at CMBP $B$-mode detection are operating at all and we
would expect this situation will continue at least for next few years. 

This paper is aimed at giving a contribution to the discussion about the first item among those listed before, that is the leading design criteria for a new instrument generation dedicated to investigate CMBP $B$-mode. The starting point will be the analysis already done to design the receivers of the SPOrt \cite{CTAL04} and BaR-SPOrt \cite{CTAL03} experiments, which represent the first attempt to realize instruments expressely oriented to measure the $Q$ and $U$ Stokes parameters. 

Section 2 will review the different main receiver architectures that can be used for CMBP experiments. Later on, Section 3 will introduce the instrument systematics, which are expected to play the major role in the race to measure the CMBP $B$-modes. Section 4 will compare the different design architectures with respect to systematics contamination and will allow us to draw some conclusions.

\section{Measuring the Polarization of the CMB}
The polarization of the CMB can be expressed by the $Q$ and $U$ Stokes parameters, which describes the linear component. The circular component V, in fact, can be considered negligible when the polarization is generated by Thomson scattering, as in the case of CMB. Even the foregrounds, Galactic Synchrotron and Dust, are considered to be only linearly polarized.  
In principle, there exist three main ways to obtain the $Q$ and $U$ Stokes parameters:
\begin{enumerate}
\item{}		By correlation of the two circularly polarized components:
    \begin{eqnarray}
        Q = \Re(E_R E_L^*); \\
        U = \Im(E_R E_L^*).
        \label{circCorrEq}
    \end{eqnarray}
It has the advantage to measure both $Q$ and $U$ simultaneously, providing thus a 100\% time efficiency.\item{}		By correlation of the two linearly polarized components:
    \begin{eqnarray}
        U = \Re(E_x E_y^*).
        \label{linCorrEq}
    \end{eqnarray} 
It is similar to the previous way, but can measure only $U$. $Q$ is obtained by rotating the reference frame of 45$^\circ$, that is by rotating the same receiver or by using another (independent) receiver. Time efficiency drops to 50\%. 
\item{}		Either by direct or off-line difference of the two linearly polarized components: 
    \begin{eqnarray}
        Q = {|E_x|^2 - |E_y|^2 \over 2}.
        \label{diffEq}
    \end{eqnarray}
    Similarly to the previous method, it provides just one parameter at a time and time efficiency is again 50\%.
\end{enumerate}
The listed situation are summarized in Table~\ref{archTab}. 
\begin{table}
 \centering
  \caption{The Stokes parameters detected and time efficiency are summarized for different architectures.}
  \begin{tabular}{@{}clcclc@{}}
    & & & & & \\
    & {\bf Architecture} & {\bf Stokes par.s} & {\bf Time Efficiency} &   &  \\
    &  {\bf detected}     &    &    &   & \\  
  \hline
    & & & & & \\
    & Correlation:						& $Q$, $U$	& 100\% &  &   \\
    & Circular Polarizations	&						&				&	 &   \\
    & & & & & \\
    & Correlation:						& $U$				& 50\%	 	&             & \\
    & Linear Polarizations		&						&					&             & \\
    & & & & & \\
    & Difference of Linear		& $Q$				& 50\%		&             & \\
    & Polarizations				 		&						& 				&             & \\
    & & & & & \\
    & Difference of Linear		& $Q$				& 50\%		&             & \\
    & Polarizations	(Off-line)&						& 				&             & \\
  \hline
  \end{tabular}
 \label{archTab}
\end{table}
An important consideration is that all of these schemes, in principle, can use both 
HEMTs and bolometric sensors. In fact, also the correlation schemes, 
which are normally thought specific for coherent (HEMT based) receivers, 
can be coupled to bolometers.
These would be used, in fact, as diodes to provide the square detection 
of the signal after the correlation units (Fig. 2). 

\begin{figure}
 \centering
  \includegraphics[angle=0, width=10cm]{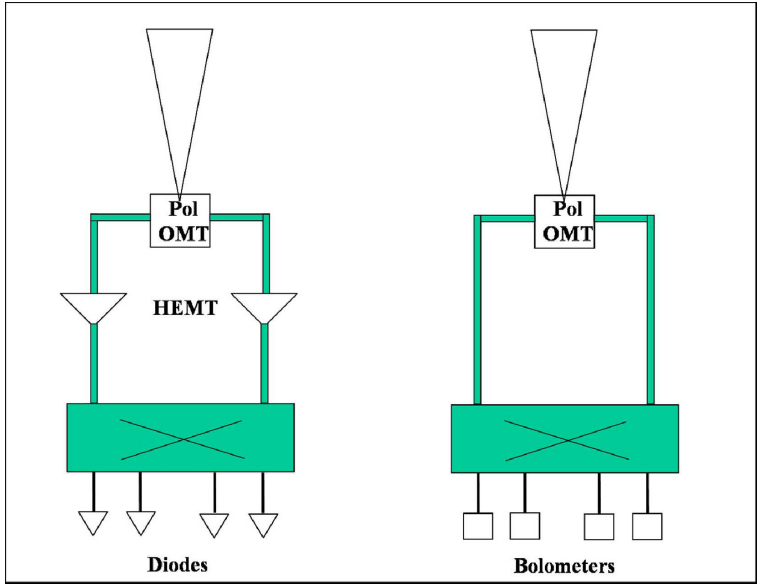}
  \caption{a correlation architecture using either HEMTs (left) or bolometers (right).} 
  \label{archFig}
\end{figure}

Figure~\ref{archFig} reports an example of how the same correlation scheme 
can adopt either HEMTs or bolometers. 
In the first case the signal is 
amplified by HEMTs, after the OMT has separated the two componets, and then is correlated by the 
correlation unit. The diodes provide square detection of the signals. 
Low noise (HEMT) amplifiers are here necessary because of the noise generated by  
diodes.

In the second case, the signal exiting the OMT is 
first correlated (without pre-amplification) and then detected 
by low 
noise bolometers, which thus act as diodes.

What has been said before has huge relevance to define baselines for designing a 
CMBP instrument. In particular we can state that: 
\begin{itemize}
\item The overall architecture can be chosen almost independently of the 
sensors to be used, leaving the designer free to optimize the architecture with 
respect to the purity of $Q$ and $U$ measurements. Any decision about the sensor type shall be driven by other criteria.
\item The sensor choice will be only marginally influenced by 
the architecture design, and other important parameters like sensitivity, 
cost, availability, overall thermal design will drive the sensor choice.
\end{itemize}
Under these assumptions, the discussion about the systematics 
generation and the purity of polarization measurements for 
the most common architectures would be mandatory.

\section{Instrument systematics and their effects in CMB Polarization measurements}

As introduced in Section 1 the measure of $B$--modes, 
for which even a 0.1\% leakage from the Temperature
anisotropy might wash out any detection, require instruments with extremely low contamination.
One should also note that the contamination from the anisotropy term
is not easily removed by destriping techniques, its level
being not constant over the scans
performed to observe the sky (e.g. see
\cite{RKA00}, \cite{SCC03} and references
therein for descriptions of destriping techniques), so that
an $intrinsically$ clean instrument is necessary to avoid complex data reductions.

A clear understanding of the contamination
from the unpolarized background introduced by
the instrument itself is thus mandatory to properly design instruments.
Several works have been written on this subject (e.g.
\cite{CTC01}\cite{LYH02}\cite{DELA02}\cite{MBB02}\cite{FFT03}\cite{PTAL03}\cite{CCS04})
and the leading effects identified so far can be divided in:
\begin{itemize}
  \item Off-axis instrumental polarization. It is generated by the optics and is independent
        of the architecture adopted for the receiver;
  \item On-axis instrumental polarization. It is generated by the receiver
        itself, in particular by the other components of the antenna assembly (e.g. OMT),
        and by the correlation devices;
  \item Systematics errors from thermal fluctuations. Are secondary systematics due
        to the fluctuations of the offset generated by the on-axis instrumental polarization.
\end{itemize}

\subsection{Optics}

The optical properties of an antenna are described by the co-polar $g(\theta,\varphi)$
 and cross-polar $\chi(\theta,\varphi)$ patterns. Both of them
 are defined for the two polarizations, so that one deals with the four
 functions $g_x$, $g_y$, $\chi_x$, $\chi_y$.
 $g$ describes the capability of the optics to collect the wanted polarization as a 
 function of the direction in the sky. For instance, $g_x(\theta, \varphi)$ describes how
 the $E_x(\theta, \varphi)$ electric field is collected into
 the polarization state $X$ of the receiver.
 
 The cross--polarization $\chi$, instead, measures how the other (wrong)
 polarization is gathered, introducing thus a contamination. 
 For example, $\chi_x$
 measures how much $E_y$ enters the polarization $X$ of the receiver. 
 
The instrumental polarization, defined as the polarization signal generated
by this device in presence of a pure unpolarized signal,
is instead described by a third antenna pattern: the instrumental
polarization pattern $\Pi$ (for details see \cite{CCS04}). 

It is a combination of $g$ and $\chi$ and its expression is given by 

\begin{equation}
    \Pi(\theta, \phi)        =
       \Pi_Q(\theta, \phi)+j \,\,\Pi_U(\theta, \phi)
       \label{piEq}
\end{equation}
with
\begin{eqnarray}
        \Pi_Q
        &=&     { |g_x|^2 + |\chi_x|^2
                 -|g_y|^2 - |\chi_y|^2
                 \over 2}, \label{fQeq} \\
        \Pi_U
        &=&     \Re\,(g_x \chi_y^* +
                     g_y \chi_x^*).
                  \label{fUeq}
\end{eqnarray}
This result depends only on the optics features and applies 
to any receiver scheme (correlation, difference between the
linear polarizations). 

\begin{figure*}
 \centering
  \includegraphics[angle=0, width=0.49\hsize]{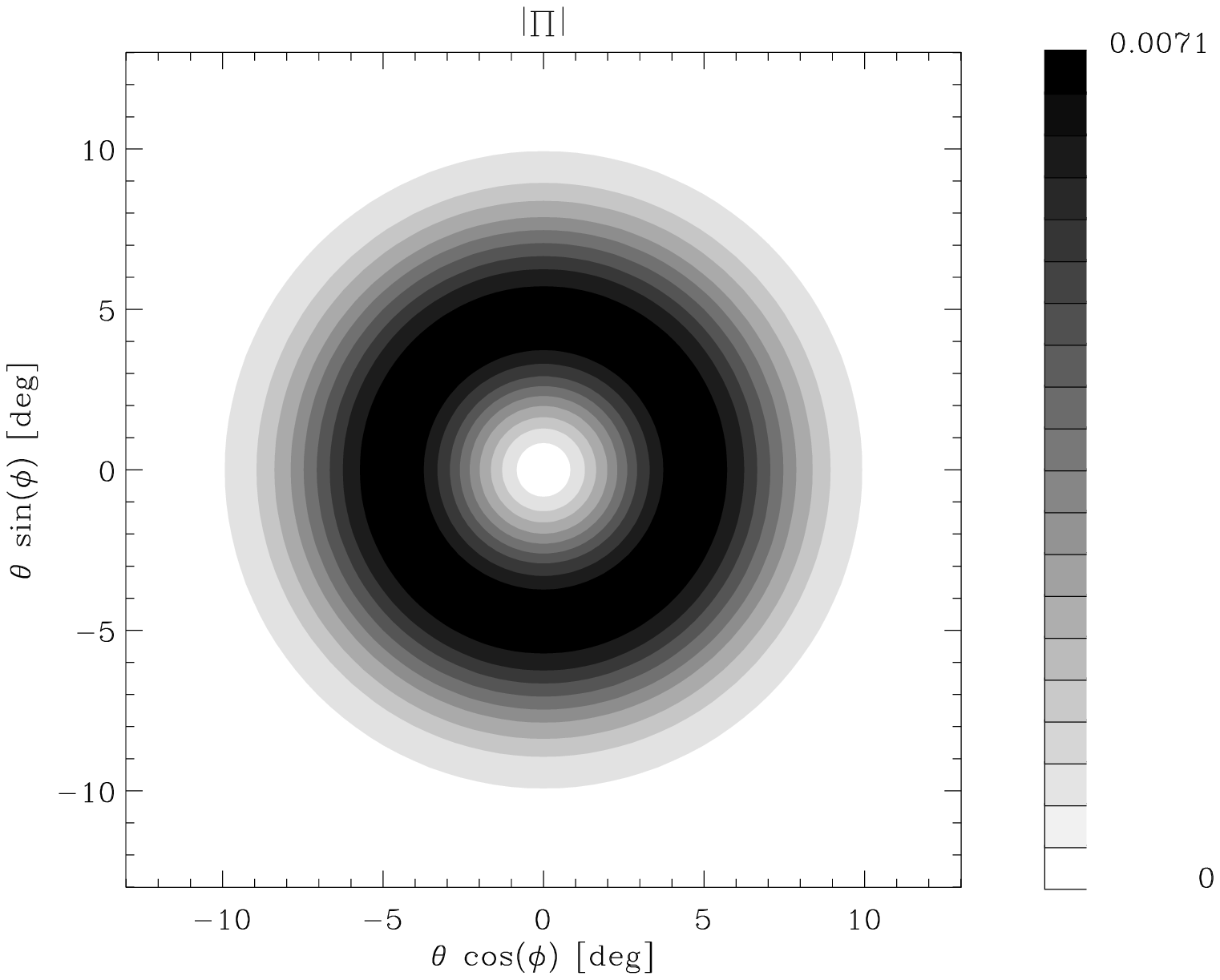}
  \includegraphics[angle=0, width=0.49\hsize]{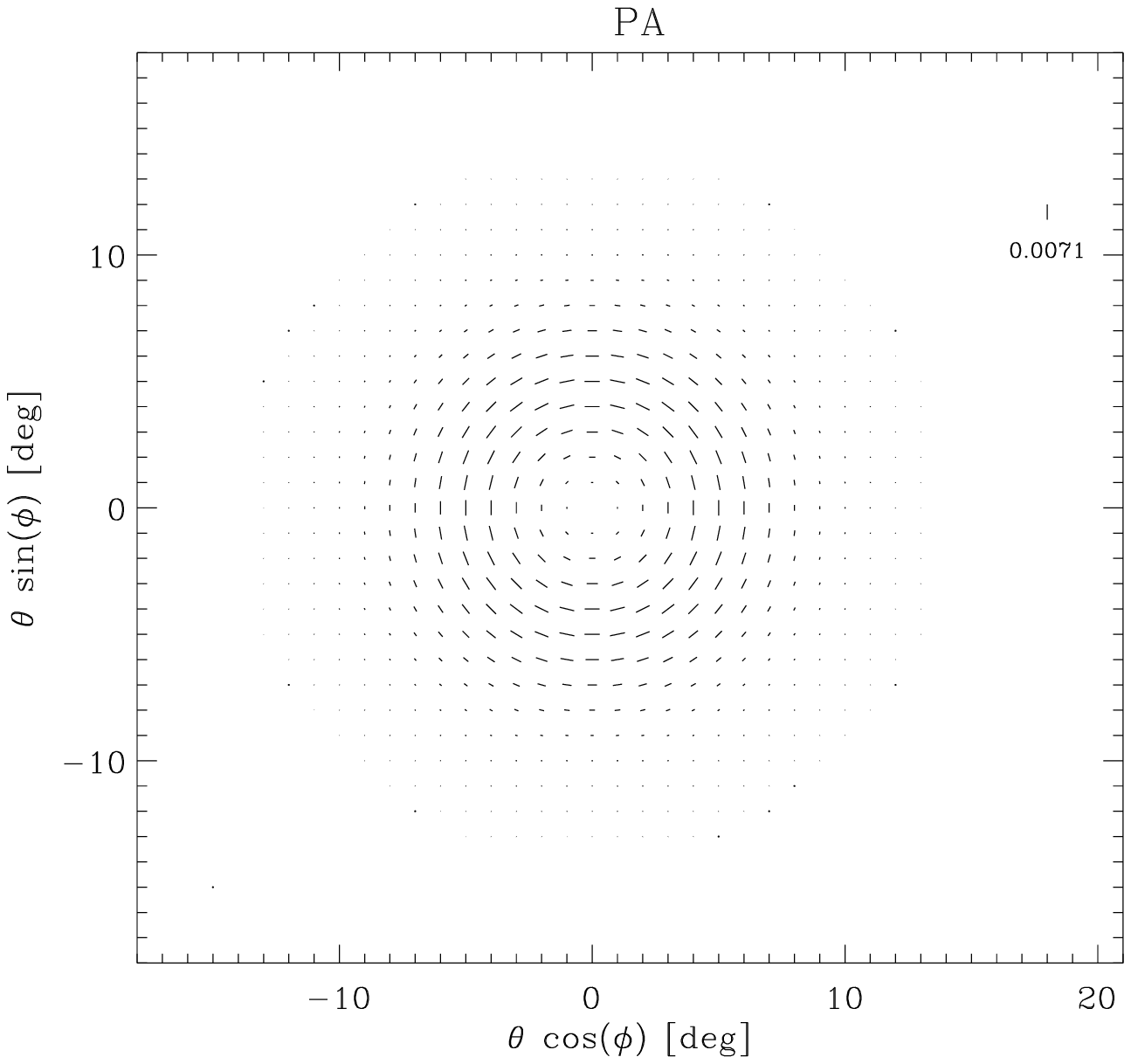}
  \includegraphics[angle=0, width=0.49\hsize]{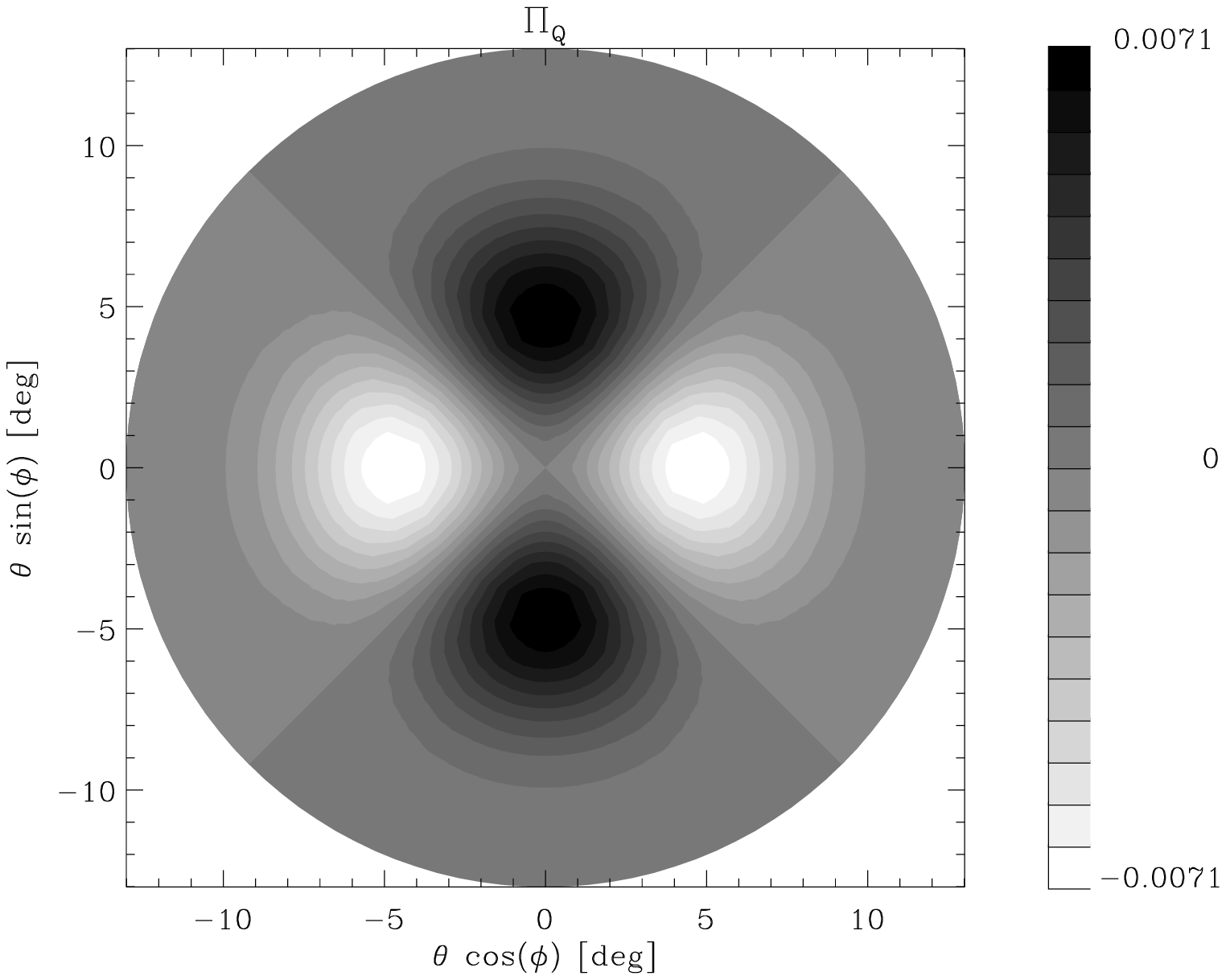}
  \includegraphics[angle=0, width=0.49\hsize]{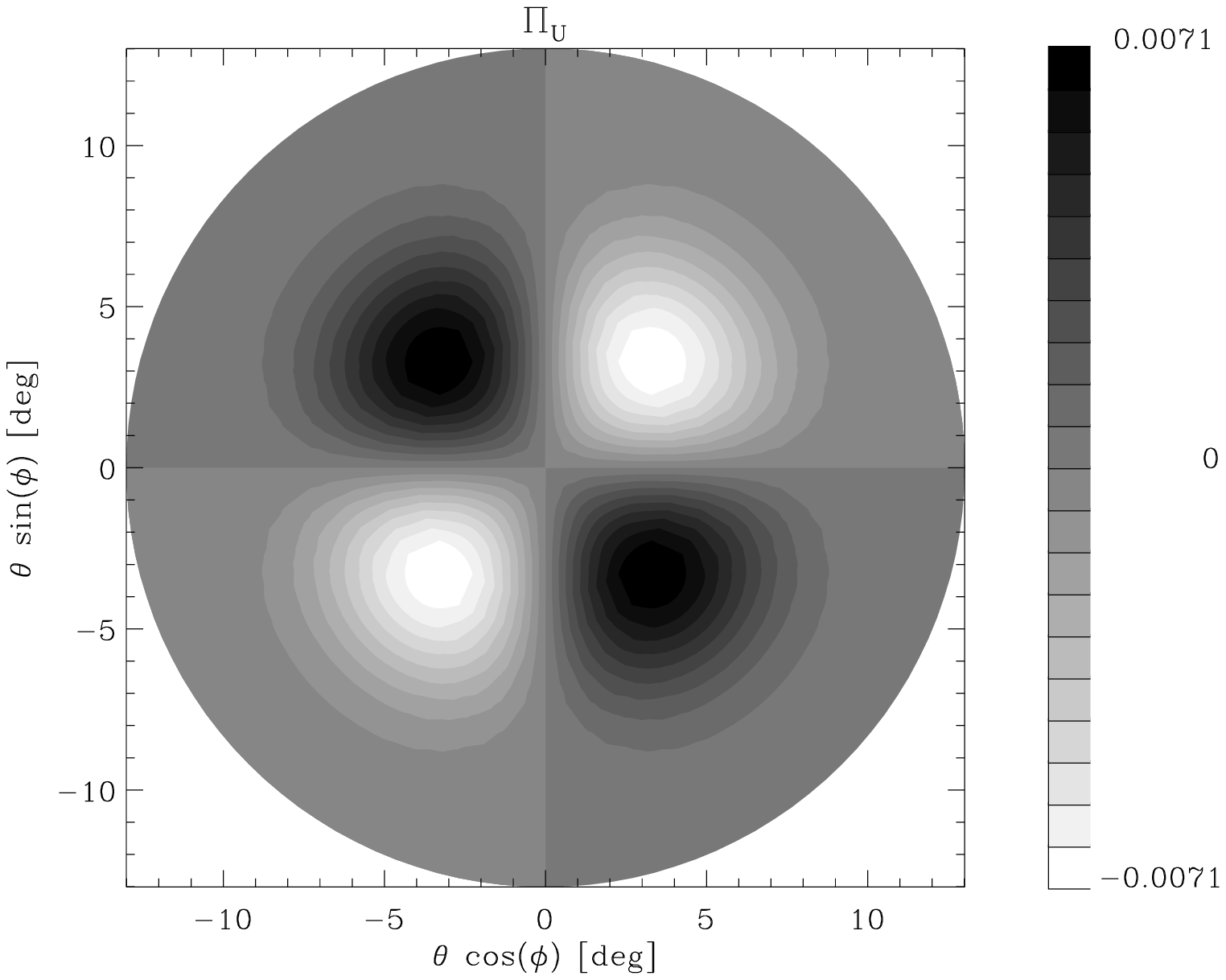}
  \caption{$\Pi$ pattern for the 90~GHz horn of the SPOrt experiment: $|\Pi|$ (top left),
           polarization angle (top right), $\Pi_Q$ (bottom left) and $\Pi_U$
           (bottom right). It is worth noting the axisymmetry of $\Pi$ and 
           the tangential pattern of the polarization angles.(From \cite{CCS04})} 
  \label{sportPIFig}
\end{figure*}

An example of this pattern is given in Figure~\ref{sportPIFig}, 
where the case of the SPOrt experiment is reported.

$\Pi$ describes just the response of the system to a single point source.
The contamination in the general case is instead given by its convolution
with the unpolarized field $T_b(\theta, \varphi)$ of the sky emission.
The equation in the antenna reference frame and in brightness
temperature unit is given by (see always \cite{CCS04})
\begin{eqnarray}
    Q_{\rm sp} + j\,U_{\rm sp} &=&
       {1 \over 4 \pi} \int_\Omega T_b(\theta, \phi) \,\, \Pi(\theta,\phi)
                 \,d\Omega. \label{tspZeq}
\end{eqnarray}
where $Q_{\rm sp}$ and $U_{\rm sp}$ are the contaminations on the two
linear Stokes parameters.

In general, evaluations of this convolution and of its effects
on CMBP measurements require a numerical approach. However,
the special case of axisymmetric antennae allows some simplifications 
which lead to an analytical solution and, in turn, help 
understand the main characteristics of such a contaminant.
Moreover, few important classes of antennae, like circular horn and
on-axis mirror optics, belong to this special case,
whose analysis has thus importance in itself.

In fact, the $\Pi$ pattern
of Equation~(\ref{piEq}) becomes
\begin{equation}
        \Pi = \Pi_Q + j\Pi_U
        = {|g_0(\theta)|^2 - |g_{\pi/2}(\theta)|^2 \over 2}\,e^{j\,2\phi}
        \label{symmfeq}
\end{equation}

where {$g_0(\theta)$ and $g_{\pi/2}(\theta)$} are the ${g_x(\theta, \phi)}$ pattern for $\theta=0$ and $\phi=\pi/2$, respectively. The intensity only depends on the angular distance $\theta$ from 
the axis and polarization angle is featured by
radial pattern with respect to the beam main axis
\begin{eqnarray}
        \alpha
        &=& 0.5 \arctan \;{U_{\rm sp} \over Q_{\rm sp}} \nonumber\\
        &=& \left\{
            \begin{array}{ll}
                 \varphi & {\rm for\,|g_0|^2 > |g_{\pi/2}|^2}\\
                 \varphi+90^\circ & {\rm otherwise}.\\
            \end{array}
        \right.
\end{eqnarray}
An example is given by the circular feed horn of 
SPOrt reported in Figure~\ref{sportPIFig}: the $\Pi$ intensity has 
axial symmetry, while the polarization angle has tangential pattern perpendicular
to the radial directions. 

First of all, we can 
observe that the maximum sensitivity is not along the centre 
of the observed field, but along an out-of-axis ring about FWHM across.
Thus, we are dealing with an off-axis instrumental polarization.

Then, $\Pi_Q$ and $\Pi_U$ have quadrilobe patterns with both
positive and negative lobes. This makes the contamination sensitive
to only the anisotropy pattern, since the contributions from the lobes and
due to the mean emission cancel out each other.
This important feature makes the contamination proportional
only to anisotropy radiation, much fainter than the mean CMB emission.

\begin{figure*}
 \centering
  \includegraphics[angle=0, width=0.49\hsize]{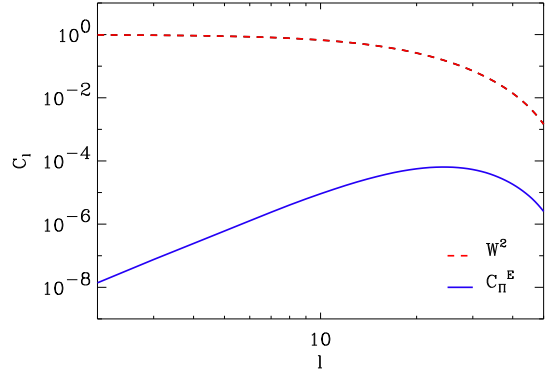}
  \caption{$E$--mode power spectrum of the $\Pi$ pattern
  for the 90~GHz feed horn of the SPOrt experiment (from \cite{CCS04}). 
  The spectrum $W_\ell$ of the
  co-polar pattern is also reported for comparison.} 
  \label{sportPIPSFig}
\end{figure*}

Finally, even more important is the radial structure of $\Pi$, which is 
typical of an $E$--mode.
This is confirmed by the computation of its polarized angular power spectra
\begin{eqnarray}
        C_{E\ell}^\Pi &=& {1\over 2\ell+1} |a_{2,\ell0}^\Pi|^2,\nonumber\\
        C_{B\ell}^\Pi &=& 0. \label{fspeceq}
\end{eqnarray}
which show no power on the $B$--mode ($a_{2,lm}^\Pi$ are the 2-spin spherical 
harmonics coefficient of $\Pi$). As an example, the $\Pi$ spectra for
the 90~GHz SPOrt antenna 
are reported in Figure~\ref{sportPIPSFig}.
It is worth noting that the $E$--mode has its maximum power
on FWHM scale, which rapidly decreases on larger angular scales: at $\ell = 2$
(90$^\circ$ scale) its value is 8 orders of magnitude lower that the 
co-polar pattern.
This interesting feature brings the maximum contamination on the smallest angular scale
accessible to the optics, while leaves clearer the largest ones.

This lack of power on the $B$--mode 
is reflected in the effects generated on the final map of the experiment.
As described by Equation~(\ref{tspZeq}), the contamination map
is the convolution between the 
Temperature map and the $\Pi$ pattern function.

An efficient way to evaluate the impact on CMBP is
the computation of the power spectrum, which represent
the most relevant quantity for CMBP purposes. 

In general it does not exist a simple analytic solution, 
and each optics configuration needs a numerical evaluation. 
This is true apart from the axisymmetric case, which
once again allows an exact solution, thus giving an idea about 
what happens also for the general case 
(see always \cite{CCS04}).
In fact, in the axisymmetric case a sort of convolution theorem is
valid and the power spectrum of the contamination
is the product between the scalar spectra of the Temperature 
map and the $E$-- and $B$--mode of the $\Pi$ pattern
\begin{eqnarray}
     C_{E\ell}^{T\otimes \Pi} &=&  {1 \over 4\pi} \, C_{E\ell}^\Pi \, C_{T\ell},\\
     C_{B\ell}^{T\otimes \Pi} &=& {1 \over 4\pi} \, C_{B\ell}^\Pi \, C_{T\ell} = 0.
     \label{cecbeq}
\end{eqnarray}

\begin{figure*}
 \centering
  \includegraphics[angle=0, width=0.49\hsize]{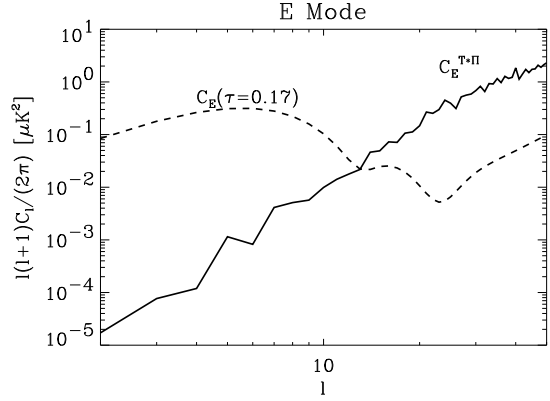}
  \caption{$E$--mode power spectrum of the contamination map
  from optics instrumental polarization for the SPOrt experiment (solid). The expected spectrum of CMBP is reported for comparison (dashed)} 
  \label{sportContPSFig}
\end{figure*}

Hence, $C_{E\ell}^\Pi$ and $C_{B\ell}^\Pi$ allow a quick evaluation of
the effects of the $T_b$ leakage: their value directly tell us 
the leakage of the $C^T$ spectrum on the polarized component. 
An example is given by Figure~\ref{sportContPSFig}, where the spectra
of the contamination map of the SPOrt case are compared to the expected CMBP 
signal.

A very important result for CMBP $B$-mode experiments is that
axisymmetric systems, having no $B$-mode component for their $\Pi$ patterns,
do not show any contamination on this CMBP component, thus representing 
the best choice for such experiments.

However, the faint CMBP signal requires many receivers in the focal plane to allow
the needed sensitivity and, apart from the central receiver, they will
be not in such an optimal condition even with an on-axis optics.

Therefore, a contamination on the $B$-mode will necessary occur and 
the optics must be carefully designed to keep under control
the leakage into this component.

However, as a representative example for $B$-mode experiments we
report a summary of the analysis carried out for the BaR-SPOrt case. The leading part of the CMBP 
$B$-mode emission is at degree scales and an experiment aimed at 
measuring it requires at least a FWHM~$= 0.5^\circ$ as angular
resolution. BaR-SPOrt, with its $0.4^\circ$ and with its
design already oriented to the CMBP signal (even if thought
for the $E$-mode only) allows a reliable evaluation of what 
is achievable with the present technology.
Actually, BaR-SPOrt has an optics with a very good cross-polarization 
(about 40--45~dB for the feed horn and much better for the mirror optics) 
and its performances provide a good
indication on what is to date achievable for instrument with
sub-degree resolution.  

\begin{figure*}
 \centering
  \includegraphics[angle=0, width=0.5\hsize]{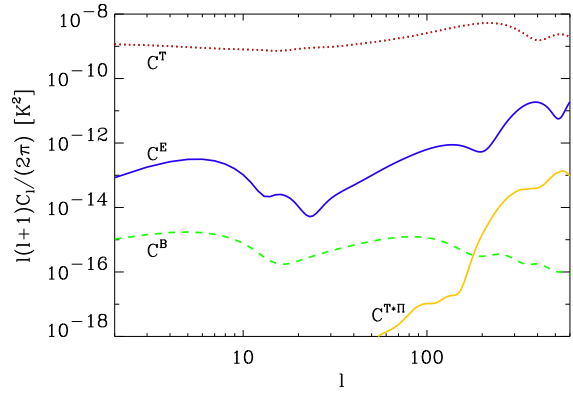}
  \caption{$E$-mode spectrum of the optics instrumental polarization
  for the BaR-SPOrt experiment ($C^{T\cdot\Pi}$). The Temperature 
  ($C^{T}$), $E$--mode ($C^{E}$) and $B$--mode ($C^{B}$) power spectra 
  of the cosmological model in agreement with the WMAP data are also
  reported for comparison. For the $B$--mode a $T/S=0.03$ ratio has been assumed.} 
  \label{barsportPSFig}
\end{figure*}

Figure~\ref{barsportPSFig} shows both the power spectra of
the contamination map. 
Having BaR-SPOrt an on-axis configuration, the results on the $B$--mode
contamination are obviously optimal (no contamination at all), but, at the
same time, this is not representative of off-axis system, for which
the $\Pi$ pattern is a mixture of both $E$ and $B$ contributions.
However, the contamination on the other component
($E$-mode) is an estimate of the whole pollution achievable, and, therefore,
it can be used as worst case indication for the contamination on the 
$B$-mode in off-axis systems, where the global contamination
is a mixture of the two modes. 

In this frame, the contamination of the optics on the $B$-mode seems to
leave free the $\ell = 100$ peak also with a low $T/S$ value, indicating that a
cross-polarization of 40--45~dB should allow us to
keep under control this contaminant. 
This source of contamination can be minimized by taking care of the optics design. 
However, this result tell us that the present technology should already have 
the right maturity level.

\subsection{Receiver}\label{recInstPolSec}

The signal collected through the antenna is elaborated and 
detected by the receiver. Sometime, as in the case of HEMT-based receivers, is amplified before detection. Since the receiver must provide the Stokes
parameters $Q$ and $U$, it should have zero output when the signal entering
the antenna is unpolarized. 
In real cases, however, the receiver introduces some 
instrumental correlation between the two polarizations, making non-zero outputs.

Differently from the optics, this part of the instrument
deals with the total power collected from the antenna independently of the incoming
direction, thus generating an on-axis instrumental polarization.
This spurious signal acts as an offset and can be seen as a correlation 
of the unpolarized signal generated by the receiver itself.

Although this offset, if constant, can be easily removed by means of
destriping techniques, it can increase 
gain instabilities of the  1/$f$--noise of detectors. 

In fact, the sensitivity radiometer equation writes
\begin{eqnarray}
  \sigma &=& T_{\rm sys}\; \sqrt{
             {k \over \Delta\nu\;\tau} + 
             \left({T_{\rm offset} \over T_{\rm sys}}\right)^2
             \left({\Delta G \over G}\right)^2
                              }
\end{eqnarray}
where $\Delta\nu$ is the bandwidth, $\tau$ the integration time,
$T_{\rm sys}$ the noise system temperature, $T_{\rm offset}$ the offset, $\Delta G / G$ the gain fluctuations and $k$ the receiver-dependent constant (e.g. $k$=1 and $k$=1/2 for total power and correlation receivers, respectively).
The first term is the white noise, which, decreasing with long 
integration times, represents the ideal behaviour of the receiver noise. 

The second term, instead, is due to the gain fluctuations,
which, worsening the ideal behaviour, can jeopardize the advantage
of long integration times.
Driven by the offset, it requires the instrumental polarization
to be as low as possible.

The total noise properties are better described by its power spectrum
which writes
\begin{eqnarray}
   P(f) &=& \sigma_0^2\;\left[ 1 + \left({f_{\rm knee} \over f}\right)^\alpha\right].
\end{eqnarray}
where $f$ is the frequency and $\sigma_0^2$ is the 1-second sensitivity. 
The first term is constant and represents the white noise. The second
term is the contribution of the gain fluctuations and, following a power law
with index $\alpha \sim$ 1 \cite{WOL95}, is called 1/$f$ noise.
The most important parameters is the knee frequency $f_{\rm knee}$. It is 
the frequency at which the 1/$f$ noise equals the white one and, thus,
defines the time scale on which the gain fluctuations becomes dominant, making
unuseful longer integration times.
Destriping techniques are typically able to remove instabilities on time scales
larger than $1/f_{\rm knee}$, which thus represents a time limit for the duration
of the scans covering the observed sky area. 

Gain instabilities are typical of active devices like HEMTs and bolometers.
As already mentioned above, 
 a way to minimize the gain fluctuations is to reduce the receiver offset. 
In fact, $f_{\rm knee}$ of a receiver is related with that of its amplifiers 
($f_{\rm knee}^{\rm LNA}$) by the formula
\begin{eqnarray}
    f_{\rm knee} &=& \left({T_{\rm offset}\over T_{\rm sys}}\right)^{2/\alpha}\, 
                     f_{\rm knee}^{\rm LNA}
                     \label{fkneeEq}
\end{eqnarray}
The present InP technology provides HEMTs with high sensitivity,
but bad stability: typical values at 100~GHz are around 
\begin{equation}
        f_{\rm knee}^{\rm LNA}\sim 10^3\;{\rm Hz.}
\end{equation}
These high values requires receivers with very low offset generation to
be compatible with the $\sim 100$~s scanning time of space experiments 
(see PLANCK and WMAP). Considering Equation~(\ref{fkneeEq}), a ratio better than
\begin{eqnarray}
   {T_{\rm offset}\over T_{\rm sys}} < 3\times10^{-3}
                     \label{totsysEq}
\end{eqnarray}
is recommended.

Better situation occurs for bolometers.
They have a typical knee frequency of \cite{DELA98}
\begin{equation}
        f_{\rm knee}^{\rm bol}\sim 10^{-2}\;{\rm Hz.}
\end{equation}
which already ensures the needed stability to keep under control 
the 1/$f$ instabilities, even in the worst case of total power receivers
where $T_{\rm offset} = T_{\rm sys}$.

Constant offsets, like those from the constant 2.7~K CMB signal or
those from the system noise, 
have impact just on the sensitivity, but do not provide
contaminations in the $Q$ and $U$ anisotropy, the relevant information
for CMBP investigations.

However, by its own definition, the offset is proportional to the 
sky emission, so that the anisotropy can generate a variable contamination
which is not trivially removable by destriping techniques.

Thanks to its on-axis nature, this effect can be described by a transfer
function $S\!P$ defined as the ratio between the instrumental polarization
and the input unpolarized signal
\begin{equation}
        S\!P = {Q_{\rm sp}+jU_{\rm sp} \over T}.
\end{equation}

To avoid to confuse these two effects, here we call {\it offset} the term related
to a constant unpolarized background (e.g. the isotropic part of CMB, the instrumental
noise) and {\it instrumental polarization} the term related to the variable
anisotropy sky emission $\Delta T$.

Considering the level of the $B$-mode component with
respect to the CMB anisotropy, the leakage from $\Delta T$ 
should be at least of the order of 
\begin{equation}
        |S\!P| \sim 10^{-3}
        \label{spCondEq}
\end{equation}
to ensure a clean detection (also in combination with off-line data reduction).

By their own nature, these contaminations can be generated only in the parts where 
the two polarizations propagates together, i.e. the antenna system and the correlation unit. 
Let us discuss only about the antenna system, which
represents the most contaminant
part. In fact, as discussed in \cite{CTC01}, 
the CU can be inserted in lock-in ring, which, introducing a modulation
onto the signal, can reject the offset generated by the CU with high efficiency.

\begin{figure}
 \centering
  \includegraphics[angle=0, width=0.6\hsize]{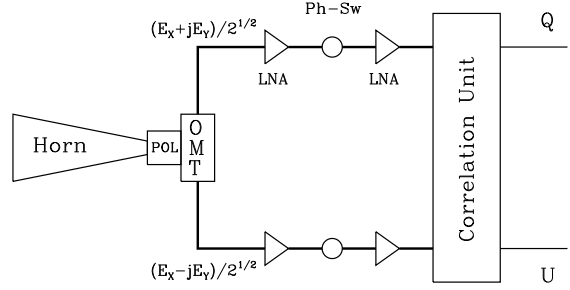}
  \caption{Scheme of polarimeter based on the correlation of the two
  circular polarizations.} 
  \label{radioFig}
\end{figure}

A first type of polarimeter is that based on the correlation of the two circularly polarized components (Figure ~\ref{radioFig}). In this architecture a polarizer is introduced after the horn to
insert a $90^\circ$ phase delay between the 
two linear polarizations $E_x$ and $E_y$ collected by the horn itself.
After that, an Orthomode Transducer (OMT) $45^\circ$ rotated extracts the two circular
polarized components
\begin{eqnarray}
    E_R &=& {E_x + jE_y \over \sqrt{2}} \nonumber \\
    E_L &=& {E_x - jE_y \over \sqrt{2}}.
    \label{erelEq}
\end{eqnarray}
The offset produced in the antenna system is due to the polarizer and the OMT,
through the expression (Carretti et al. 2001)
\begin{eqnarray}
 (Q+jU)_{\rm offset}  
             &=& S\!P_{\rm omt}
               \left(T_{\rm sky} + T_{\rm atm} +
                 T_{\rm noise}^{\rm Ant}
               \right)  \nonumber \\
             &+&   S\!P_{\rm pol}
                  \left(T_{\rm sky} + T_{\rm atm} +
                            T_{\rm noise}^{\rm h} -
                            A_{\rm h}\;T_{\rm ph}^{\rm p}
                          \right),\nonumber \\
                  \label{AB0TN0eq}
\end{eqnarray}
where $S\!P_{\rm omt}$ and $S\!P_{\rm pol}$ are
\begin{eqnarray}
 S\!P_{\rm omt} & = & A_{\rm omt}\;\left(S_{A1}S_{B1}^* + S_{A2}S_{B2}^*\right),
                     \label{spomtEq} \\
                     \nonumber \\
 S\!P_{\rm pol} & = & {1\over 2} \left(1 - {A_{\parallel}\over
                                 A_{\perp}}\right) 
                  =  {1\over 2} \; {A_{\perp} - A_{\parallel}
                       \over A_{\perp}},
\label{sppolEq}
\end{eqnarray}
$S_{A1}$, $S_{B2}$ the transmission parameters of the two OMT arms,
$S_{A2}$, $S_{B1}$ their isolation terms, $A_{\rm h}$, $A_{\rm omt}$
the attenuations of the horn and the OMT, respectively, 
and $A_{\perp}$, $A_{\parallel}$ the attenuations 
of the polarizer along its two main polarizations.
The offset sources are thus
the physical temperature of the polarizer 
$T_{\rm ph}^{\rm p}$ and the signals propagating in the antenna system:
the signal collected from the sky ($T_{\rm sky}$), the 
atmosphere emission ($T_{\rm atm}$) and the noise temperatures
of the horn alone ($T_{\rm noise}^{\rm h}$) 
and of the whole antenna system
\begin{equation}
  T_{\rm noise}^{\rm Ant} = T_{\rm noise}^{\rm h} +
  A_{\rm h}\; T_{\rm noise}^{\rm p} + 
  A_{\rm h}\;A_{\rm p}\; T_{\rm noise}^{\rm omt},
  \label{tantnoiseeq}
\end{equation}
with $A_{\rm p}$ the mean attenuation of the two polarizer arms and
\begin{eqnarray} 
T_{\rm noise}^{\rm h}  & = &                           
               (A_{\rm h} - 1)\;T_{\rm ph}^{\rm h}\label{thneq}\\
T_{\rm noise}^{\rm p}   & = &   
               (A_{\rm p } - 1)\;T_{\rm ph}^{\rm p}\label{tpneq}\\
T_{\rm noise}^{\rm omt}   & = &
               (A_{\rm omt } - 1)\;T_{\rm ph}^{\rm omt}\label{toneq}
\end{eqnarray}
with $T_{\rm ph}^{\rm h}$, $T_{\rm ph}^{\rm p}$, $T_{\rm ph}^{\rm omt}$
the physical temperatures of horn, polarizer and OMT.

A complete description of derivation
and implications of Eq.~(\ref{AB0TN0eq}) is given in
Carretti et al. (2001). Here we 
just point out that the offset is generated by OMT and polarizer
that partially correlate the antenna noise
as well as the sky and atmosphere emissions. 

The part of Equation~(\ref{AB0TN0eq}) concerning $T_{\rm sky}$ provides the instrumental
polarization related to the CMB anisotropy:
\begin{eqnarray}
 (Q+jU)_{\rm sp}  &=& \left( S\!P_{\rm omt} +   S\!P_{\rm pol}\right) \; \Delta T_{\rm sky}
                  \label{quspDTEq}
\end{eqnarray}
so that $S\!P_{\rm omt}$ and $S\!P_{\rm pol}$ directly provides the leakage 
$S\!P$ of $\Delta T_{\rm sky}$ into $Q+jU$.

A second class of polarimeters are the correlators of the two linear polarizations.
The antenna system is very similar to the previous one, 
but the polarizer, which is not inserted in the chain. 
The OMT, thus, directly extract the two linear polarizations which are correlated
providing $U$ only, as for Equation~(\ref{linCorrEq}).

Apart from the lack of the polarizer term, 
the expression describing the offset is very similar to equation~(\ref{AB0TN0eq}) and
writes
\begin{eqnarray}
 Q_{\rm sp}+jU_{\rm sp}  
             &=& S\!P_{\rm omt}
               \left(T_{\rm sky} + T_{\rm atm} +
                 T_{\rm noise}^{\rm Ant}
               \right).
               \label{linOffEq}
\end{eqnarray}
Similarly, the leakage from CMB anisotropy into the polarized signal is
\begin{eqnarray}
 (Q+jU)_{\rm sp}  &=& S\!P_{\rm omt}\;\Delta T.
                  \label{linQuspDTEq}
\end{eqnarray}

In general, the major term is that concerning the OMT,
so that the lack of the polarizer does not 
give relevant differences between the two correlation
schemes in terms of offset  and instrumental polarization generation.

The third polarimeter scheme consists in a differential receiver subtracting
the two linear polarizations and providing directly $Q$.
Discussed by \cite{LYH02}, it is realized by a dual polarization 
antenna directly feeding an OMT, whose outputs are $E_x$ and $E_y$. 
They are then differentiated 
by a PLANCK-like receiver, with the second polarization which substitutes
the 4~K reference load: $E_x$ and $E_y$ enter a magic-T providing
their sum and difference ($(E_x + E_y)/\sqrt{2}$ and $(E_x - E_y)/\sqrt{2}$).
After the amplification, these enter a second magic-T which provide two 
quantities proportional to $E_x$ and $E_y$. After a square detection performed
by diodes, they are differentiated thus providing $Q$.

The offset generated by the antenna system is given by
\begin{eqnarray}
 (Q+jU)_{\rm offset}  
             &=& S\!P_{\rm omt}^{\rm diff}
                  \left(T_{\rm sky} + T_{\rm atm} +
                            T_{\rm noise}^{\rm h} -
                            A_{\rm h}\;T_{\rm ph}^{\rm omt}
                          \right),
                  \label{diffOffEqeq}
\end{eqnarray}
with 
\begin{eqnarray}
     S\!P_{\rm omt}^{\rm diff}&=&  
                     {1\over 2} \; {A_{Y}^{\rm omt} - A_{X}^{\rm omt}
                       \over A^{\rm omt}},
\end{eqnarray}
where $T_{\rm ph}^{\rm omt}$ is the physical temperature of the OMT,
$A_{X}^{\rm omt}$ and $A_{Y}^{\rm omt}$ the attenuations
of the two OMT arms and $A^{\rm omt}$ the mean OMT attenuation.
Therefore, in this case the differential attenuation of the OMT, instead of
the polarizer, is the main cause of offset generation.

Similarly, the instrumental polarization proportional to the CMB anisotropy is
given by 
\begin{eqnarray}
 (Q+jU)_{\rm sp}  
             &=& S\!P_{\rm omt}^{\rm diff}\;\Delta T_{\rm sky}
                  \label{diffspEqeq}
\end{eqnarray}
Finally, polarization can be measured also with a pure bolometric 
technique. 
Polarization Sensitive Bolometers (PSBs) are able to separately measure the
power of the two linear polarizations ($|E_x|^2$ and $|E_y|^2$). 
The Stokes parameter $Q$ is computed by differencing the power of the
two polarizations so detected. The Stokes parameter $U$ cannot be directly measured 
and, as for differential polarimeters, have to be captured through a second
measurement $45^\circ$ rotated.

As OMTs do for coherent receivers, there is leakage between 
the two polarizations. 
Following \cite{JON02} this leakage is described through $\epsilon_x$, 
the fraction of power of $|E_y|^2$ detected in the $|E_x|^2$ channel, 
and through $\epsilon_y$, the equivalent for the other polarization. 
The instrumental polarization from anisotropy signal is thus given by 
\cite{JON02}
\begin{eqnarray}
 Q_{\rm sp}^{\rm psb} &=& SP_{\rm psb} \Delta T_{\rm sky},
  \label{psbInst1Eq}
\end{eqnarray}
with
\begin{eqnarray}
 SP_{\rm psb} &=& {\epsilon_x - \epsilon_y \over 2},
  \label{psbInst2Eq}
\end{eqnarray}
which provides the way to measure the non ideality of PSB detectors.

\subsection{Thermal fluctuations}
Thermal fluctuations are widely recognized as a possible source of errors in faint CMB measurements \cite{PTAL03}\cite{MBB02}\cite{CZM04}\cite{PIAT03} and, besides a quiet environment, receivers with 
low sensitivity to temperature
are required for the detection of signal as weak as CMBP.  

The importance of such systematics depends on the receiver scheme and,
in some cases, can be crucial. 
Total power architectures suffer of variations induced
onto the data, that are only slightly dumped with respect to the 
temperature fluctuations themselves. The detection
of a weak ($\mu$K) signal is thus a real challenge 
even in very stable thermal environments.

Carretti et al. (2004b) describes such effects in terms of transfer functions 
from physical temperature variations to fluctuations onto the data. 
In fact, these transfer functions allow a direct 
estimate of the sensitivity to temperature fluctuations of the 
different architectures and, in turn, a direct comparison among these schemes.

For correlation receivers like those of SPOrt and BaR-SPOrt, they found that the
variations on the data are generated by the thermal fluctuations 
$\Delta T_{\rm ph}^{\rm h}$, $\Delta T_{\rm ph}^{\rm p}$, 
$\Delta T_{\rm ph}^{\rm omt}$ of OMT, polarizer and feed horn, respectively,
and are described by the equation
\begin{eqnarray}
 \Delta(Q+jU)   & = &
           H_{\rm h}\; \Delta T_{\rm ph}^{\rm h}\nonumber \\
			 &+& H_{\rm p}\;  \Delta T_{\rm ph}^{\rm p}\nonumber  \\
			 &+& H_{\rm omt}\;  \Delta T_{\rm ph}^{\rm omt}
		  \label{offFluct1eq}
\end{eqnarray}
where the transfer functions $H$ of the three devices are defined by
\begin{eqnarray}
          H_{\rm h}   &=& {3\over 2}\,(A_{\rm h} - 1)
     \;[S\!P_{\rm omt}(1+h^{\rm omt}) + S\!P_{\rm pol}(1+h^{\rm p})],\nonumber\\
          \nonumber\\
          H_{\rm p}   &=& {3\over 2}\,A_{\rm h}\;[S\!P_{\rm omt}\,(A_{\rm p} - 1)\, 
                                     (1+p^{\rm omt})-
			                      S\!P_{\rm pol}(1+p^{\rm p})],		\nonumber \\
          \nonumber\\
			    H_{\rm omt} &=& {3\over 2}\,A_{\rm h}\;A_{\rm p}\;(A_{\rm omt} - 1)
			                     \;S\!P_{\rm omt}\;(1+O^{\rm omt}),
			   \label{hhheq}
\end{eqnarray}
for horn (h), polarizer (p), and OMT (omt), respectively. $A$ are
the attenuations of the devices, while the corrective 
terms $h$, $p$, $O$ are
\begin{eqnarray}
          \nonumber\\
          h^{\rm omt}  &=& (A_{\rm p} - 1)\,{T_{\rm ph}^{\rm p}\over
                            3\;T_{\rm ph}^{\rm h}}
                          +A_{\rm p}\,(A_{\rm omt} - 1)\,{T_{\rm ph}^{\rm omt}\over
                            3\;T_{\rm ph}^{\rm h}},\nonumber\\
          h^{\rm p}  &=& -{T_{\rm ph}^{\rm p},\over
                            3\;T_{\rm ph}^{\rm h}},\nonumber\\
          p^{\rm omt}  &=& (A_{\rm omt}-1)\,{T_{\rm ph}^{\rm omt}\over
                            3\;T_{\rm ph}^{\rm p}},\nonumber\\
          p^{\rm p}  &=& -\left(
                            {A_{\rm h}-1 \over A_{\rm h}}\, {T_{\rm ph}^{\rm h}\over
                             3\;T_{\rm ph}^{\rm p}}
                            + {1 \over A_{\rm h}}\, {T_{\rm sky}+T_{\rm atm}\over
                             3\;T_{\rm ph}^{\rm p}}
                          \right), \nonumber\\
          O^{\rm omt}  &=& {A_{\rm h}-1 \over A_{\rm h}\,A_{\rm p}}\, 
                               {T_{\rm ph}^{\rm h}\over 6\;T_{\rm ph}^{\rm omt}}
                        +  {A_{\rm p}-1 \over A_{\rm p}} \,
                               {T_{\rm ph}^{\rm p}\over 6\;T_{\rm ph}^{\rm omt}}
                        +  {A_{\rm omt}-1 \over 6} \nonumber\\
                       & &+  {1 \over A_{\rm h}\,A_{\rm p}}\, 
                               {T_{\rm sky}+T_{\rm atm}\over 6\;T_{\rm ph}^{\rm omt}},
                               \nonumber\\
 			   \label{hpOeq}
\end{eqnarray}
which, dominated by $(A-1)$ terms, 
are in general much lower than 1.

The dumping factors with respect to the thermal fluctuations
are thus given by noise generation terms ($A-1$) and by
the extra-terms ($S\!P_{\rm omt}$, $S\!P_{\rm pol}$) typical of the correlation 
architecture. 

The total power scheme, instead, does not benefit of such extra-factors. 
Actually, transfer functions for this scheme are \cite{CZM04}
\begin{eqnarray}
          H_{\rm h}^{TP} &=& {3\over 2}\;(A_{\rm h} - 1)\;(1+h^{\rm omt}),\\
          H_{\rm p}^{TP}  &=& {3\over 2}\;A_{\rm h}\;(A_{\rm p} - 1)\;(1+p^{\rm omt}),		 \\
			    H_{\rm omt}^{TP}&=& {3\over 2}\;A_{\rm h}\;A_{\rm p}\;(A_{\rm omt} - 1).
			   \label{hhhTPeq}
\end{eqnarray}
Considering that $S\!P_{\rm omt}$ and $S\!P_{\rm pol}$ can be as low as 
$10^{-3}$ (see Section~\ref{recInstPolSec}),
the advantages of correlation architectures become obvious.

Beside the transfer functions, which describe the capability of a scheme
to be insensitive to temperature fluctuations, the contamination
on the final map of a CMB experiment have to be evaluated to check
the importance with respect to the CMB signal.
The effects on the final map (and on their spectra, the relevant quantities
for CMB studies) are too complex to be analytically estimated: they
depend on the amplitude of the thermal fluctuations,
their statistical behaviour and, finally, the scanning strategy of the experiment.
Simulations are thus needed and a general rule cannot be provided:
estimates have to be performed for every single experiment.

However, as an example, we can consider the SPOrt case, which, on board the 
International Space Station, will not benefit of optimal thermal
conditions. In fact, the ISS orbit is featured by 
a daily modulation of the Sun illumination generating
significative environmental temperature fluctuations.
Therefore, it represents a sort of worst case analysis of the
performances achievable by an experiment devoted to the CMBP $B$--mode. 
The best case would be a correlation architecture SPOrt-like (see next section) coupled with an orbit characterized by optimal thermal environment (e.g. L2 point of the Sun--Earth system). 
The analysis of \cite{CZM04} shows that, although the non-optimal
thermal environment, the contamination from thermal fluctuations are at 
a low level for this experiment.
In fact, the correlation scheme,
along with the custom developed devices (OMT and polarizer),
allows very small transfer functions
(about $5\times 10^{-4}$ for the worst device). In combination with an active control
of the temperature fluctuations, which reduces to $\pm 0.2$~K the 
{\it natural} thermal fluctuations induced by the Sun modulation,
this allows to keep under control the contamination from thermal instabilities.
Figure 6 in \cite{CZM04} shows the thermal induced noise in comparison with
the expected $E$--mode spectra, main scientific target of the SPOrt experiment:
the contamination is well below the sky signal at the multipole $\ell$ where
it provides the major contribution. 

A further point is the relevance of the thermal fluctuation behaviour 
\cite{MBB02}\cite{CZM04}. In fact, fluctuations synchronous
with the scanning period generate the major contribution. The thermal design has
thus to be optimized to minimize this component. This work has been carried out
on both the Planck and SPOrt experiments.

The case of the SPOrt experiment shows that a good thermal design, in combination
with a receiver generating very low offsets, allows to keep under control the
thermal fluctuation effects even in non-optimal orbits.
Even better results are expected for SPOrt-like instruments in 
orbits which are optimal from the thermal stability point of view.

Therefore, this kind of systematics can be minimized by a careful instrument thermal design and by a mission design able to reduce the orbit synchronous thermal fluctuations.

\section{Architectures vs Systematics contamination}

From the point of view of the instrumental polarization by optics, all the architectures are 
equivalent, the contamination depending on the optics features only.
Optics design must be carefully studied, but the analysis of the 
BaR-SPOrt case gives indications that present technology can provide antennae free enough from this
contaminant, even for the $B$--mode detection.

The other relevant source of contamination is the instrumental
polarization by the receiver. In this case there are relevant differences among the various
schemes.
The purity can be evaluated through only the $S\!P$ coefficient giving
the instrumental polarization from CMB anisotropy (Section 3.2). This is because the 
offset generation and related 1/$f$ noise effects are in general less critical.
In fact, both the offset and instrumental polarization are generated by the 
same coefficient, but the condition to be matched for the instrumental polarization
(Equation~(\ref{spCondEq})) is more severe than 
that for the offset (Equation~(\ref{totsysEq})). 

Table~\ref{coeffTab} provides the coefficient $S\!P$
for the different architectures discussed here. 
The computations have been performed assuming the best performances available to date. In particular, for the correlation polarimeters we have assumed the
$S\!P$ values of the SPOrt/BaR-SPOrt experiments, 
which presently represent the state-of-the-art of such a scheme in terms of
$Q$ and $U$ purity. For these experiments $S\!P\sim 2 \times 10^{-3}$ (90 GHz channel of SPOrt) represents the worst case, while a better $S\!P\sim 2 \times 10^{-4}$ has been obtained for the 32 GHz channel of BaR-SPOrt.
\begin{table}
 \centering
  \caption{Instrumental polarization coefficient $S\!P$ for various polarimeter
           schemes (brackets report values achieved at 32~GHz which can 
           be taken as goals for future developments).}
  \begin{tabular}{@{}clcc@{}}
  \\
  & {\bf Architecture} & $S\!P$ &\\
  \hline \\
  & Correlation:						& $\sim 2 \times 10^{-3}$ & \\
  & Circular Polarizations	&	 ($2\times 10^{-4}$ at 32~GHz) & \\
  & & & \\
  & Correlation:						& see circular polarizations& \\
  & Linear Polarizations		&	& \\
  & & & \\
  & Difference of Linear		& $\sim 1 \times 10^{-2}$ & \\
  & Polarizations				 		&	 ($3\times10^{-3}$ at 30~GHz)& \\
  & & & \\
  & Difference of Linear		& & \\
  & Polarizations	(Off-line)&	$\sim 1. \times 10^{-2}$ & \\
  & and PSB                 &	& \\
  \hline
  \end{tabular}
 \label{coeffTab}
\end{table}

For the differential scheme, instead, we have assumed a differential attenuation of
$0.1~dB$. This is the performance of
WMAP 94~GHz channel \cite{JAR03}, 
but it should be noted that the 30~GHz SPOrt OMTs allow better results.

Finally, typical values for PSB leakage are $\epsilon_i < 0.05$ \cite{JON02}, 
about 4-5 orders of magnitude worst than the
best OMTs, leading instrumental polarization coefficient of the order of
$SP_{\rm psb} \sim 1\times 10^{-2}$, similar to that of the differential
receiver.

\section{Conclusions}

The above discussion brings to the fore that there is a scheme already matching (or very close to) the 
level of purity required for a $B$--mode experiment ($S\!P \cong 10^{-3}$): the correlation scheme of the SPOrt and BaR-SPOrt experiments (see Table 2).
That scheme was studied taking into account the major requirement of \textit{high-purity} for CMBP measurements and the sources generating 
contamination have been deeply analyzed.
A design-to-goal philosophy has been followed by developing the most critical
components. Then, a custom approach based on the optimization of the
devices with respect to the instrumental polarization generation has brought to this important result.

Even though the correlation of the two circular polarizations 
may already satisfies the requirements at least in some cases, other schemes
like the differential ones could have margin for improvements. 
Further developments, in fact, should allow
the improvement of the $\approx 1$ order of magnitude 
required by differential architectures to get to the same result.

However, it is worth remarking that only the correlation of the two circular polarizations allows 
simultaneous detection of the two Stokes parameters $Q$ and $U$, which offers  two unvaluable advantages with respect to other schemes:
\begin{itemize}
   \item the time efficiency is a factor 2 better than for the others, allowing either a $\sqrt{2}$ better sensitivity or, alternatively,
      half a number of receivers. In turn, lower costs and system resources (mass,
      power) are realistic;
   \item{} $Q$ and $U$ are detected by the same receiver, allowing
           to avoid the complex coupling of data coming from
           different instruments, and more complex calibration procedure.
\end{itemize}

In the light of this, the design of future CMBP experiments aiming at $B$--modes detection could start by taking the overall architecture of SPOrt as a baseline design for developing arrays of receivers, in order to achieve the needed sensitivity. Thus, the capability of realizing arrays of hundred receivers with suitable polarimetric performances will represent the major challenge for next generation of instruments. 

It is worthwhile noting that SPOrt-like architectures are suitable to be used even with bolometers, taking advantage of their higher sensitivity. In this case the signal amplification would not be required before detection, then avoiding additional noise introduced by front-end amplifiers.
However, this solution implies more complicated cryogenics design due to the need of
cooling bolometers down to $\sim 1 K$, and even the front-end (feed, OMT, and correlation
unit) must be cryogenically cooled (down to few K) to reduce their thermal noise. In that case the
cryogenics must be performed by using liquid Helium and, apart from its intrinsic
complexity, also the time duration of the mission will be constrained.
The alternative of using Low Noise Amplifiers, before detection with diodes, does not
require the use of liquid Helium reservoir since the required temperatures 
($\approx 15$--20~K) can be provided by mechanical coolers. 

In the light of this, a really preliminary road map for future CMBP experiments \textit{B--mode--oriented} can be written down:
\begin{itemize}
	\item adopt a correlation scheme of circular components to have both $Q$ and $U$ simultaneosly
	\item develop array of receivers based on the SPOrt design for passive devices
	\item assess \textit{pro et contra} of using bolometers instead of LNA and diodes in term of impact into cryogenics design
	\item put most of efforts into technological development of either LNA or bolometers depending on the cryogenics assessment

\end{itemize}

The above list suggests the cryogenics could thus represent the watershed line. If space cryogenics able to ensure proper cooling to bolometers (and front-end) can be available soon, that is in time for incoming CMBP space missions, then the best effort shall be addressed towards bolometers in order to make it competitive at frequency around 100 GHz. On the other hand, if the sooner available space cryogenics will be that suitable for HEMT technology, then the goal shall be the realization of array of W-band LNAs with improved performances.

\end{document}